\documentclass[symmetry,article,accept,moreauthors,pdftex]{Definitions/mdpi} 

\firstpage{1} 
\pubvolume{1}
\issuenum{1}
\articlenumber{0}
\pubyear{2022}
\copyrightyear{2022}
\externaleditor{Academic Editor: 
 Firstname Lastname} 
\datereceived{} 
\dateaccepted{} 
\datepublished{} 
\hreflink{https://doi.org/} 

\newcommand{\mnras}{Mon. Not. R. Astron. Soc.}
\newcommand{\apj}{Astrophys. J.}
\newcommand{\apjs}{Astrophys. J. Suppl. Ser.}
\newcommand{\apjl}{Astrophys. J. Lett.}
\newcommand{\aj}{Astron. J.}
\newcommand{\aap}{Astron. Astrophys.}
\newcommand{\aapr}{Astron. Astrophys. Rev.}

\Title{The Quasar CTD\,135 is Not a Compact Symmetric Object}

\TitleCitation{The Quasar CTD\,135 is Not a Compact Symmetric Object}

\Author{S\'{a}ndor Frey $^{1,2,}$*\orcidA{}, Krisztina \'{E}. Gab\'{a}nyi $^{1,3,4}$\orcidB{} and Tao An $^{5}$\orcidC{} }

\AuthorNames{S\'{a}ndor Frey, Krisztina \'{E}. Gab\'{a}nyi and Tao An}

\AuthorCitation{Frey, S.; Gab\'{a}nyi, K.\'{E}.; An, T.}

\address{%
$^{1}$ \quad Konkoly Observatory, Research Centre for Astronomy and Earth Sciences, Konkoly Thege Mikl\'{o}s \'{u}t 15-17, H-1121 Budapest, Hungary;  k.gabanyi@astro.elte.hu \\
$^{2}$ \quad Institute of Physics, ELTE E\"otv\"os Lor\'and University, P\'{a}zm\'{a}ny P\'{e}ter s\'{e}t\'{a}ny 1/A, H-1117 Budapest, Hungary\\
$^{3}$ \quad Department of Astronomy, Institute of Geography and Earth Sciences, ELTE E\"{o}tv\"{o}s Lor\'{a}nd University,  P\'{a}zm\'{a}ny P\'{e}ter s\'{e}t\'{a}ny 1/A, H-1117 Budapest, Hungary\\
$^{4}$ \quad ELKH-ELTE Extragalactic Astrophysics Research Group, ELTE E\"otv\"os Lor\'and University, P\'{a}zm\'{a}ny P\'{e}ter S\'{e}t\'{a}ny 1/A, H-1117 Budapest, Hungary\\
$^{5}$ \quad Shanghai Astronomical Observatory, Key Laboratory of Radio Astronomy, Chinese Academy of Sciences, 80 Nandan Road, Shanghai 200030, China; antao@shao.ac.cn
}

\corres{Correspondence: frey.sandor@csfk.org}

\abstract{The radio-loud quasar CTD\,135 (2234+282, J2236+2828) has been proposed as a candidate compact symmetric object (CSO), based on its symmetric radio structure revealed by multi-frequency very long baseline interferometry (VLBI) imaging observations on milliarcsec angular scales. CSOs are known as young jetted active galactic nuclei (AGN) whose relativistic plasma jets are misaligned with respect to the line of sight. The peculiarity of CTD\,135 as a CSO candidate was its detection in $\gamma$-rays, while the vast majority of known $\gamma$-ray emitting AGN are blazars with jets pointing close to our viewing direction. Since only a handful of CSOs are known as $\gamma$-ray sources, the unambiguous identification of a single candidate is important for studying this rare class of objects. By collecting and interpreting observational data from the recent literature, we revisit the classification of CTD\,135. We present evidence that the object, based on its flat-spectrum radio core with high brightness temperature, variability at multiple wavebands, and infrared colours should be classified as a blazar rather than a CSO.}

\keyword{active galactic nuclei; blazars; compact symmetric objects; very long baseline interferometry; $\gamma$-rays; jets; variability}

\begin{document}

\section{Introduction}

Compact symmetric objects (CSOs) are physically small (<1\,kpc) jetted active galactic nuclei (AGN) with powerful radio emission. As~their name coined by~\cite{1994ApJ...432L..87W} indicates, they show a symmetric ``mini-lobe'' structure around a central core. The~core itself often remains weak or even undetected in high-resolution very long baseline interferometry (VLBI) imaging because the jets originating from the vicinity of the central supermassive black hole are inclined to the line of sight by a large angle. It is in sharp contrast with blazars, where jets nearly align with the viewing direction. The~synchrotron radio emission of blazar jets is enhanced by relativistic beaming, while the emission of CSOs is dominated by the shock fronts where the energetic jets interact with the ambient~material. 

When monitored with VLBI, blazars typically show apparent superluminal jet component motion, a~combined effect of a plasma outflow whose speed is intrinsically close to the speed of light, and~the small jet inclination~\cite{1995PASP..107..803U}. On~the other hand, the~apparent hotspot expansion speed in CSOs is mostly subluminal. CSOs are known to represent the earliest evolutionary phase of symmetric jetted radio AGN~\cite{1995A&A...302..317F,1998A&A...337...69O,2000ApJ...541..112T,2012ApJ...760...77A}. 

Blazars are by far the most common AGN class found among the associated extragalactic objects in the most recent $\gamma$-ray source catalogue of the \textit{Fermi} Large Area Telescope (LAT)~\cite{2020ApJS..247...33A}. However, there are theoretical works predicting $\gamma$-ray emission also from CSOs~\cite{2008ApJ...680..911S,2009MNRAS.395L..43K,2014ApJ...780..165M}. Only a handful of CSOs and somewhat more extended radio AGNs (e.g.,~\cite{2017MNRAS.466..952A}) with misaligned jets are known as $\gamma$-ray sources (see a recent review in~\cite{2021A&ARv..29....3O} and references therein). The radiation mechanism and the exact location where the $\gamma$-rays come from are still under debate. Unlike for blazars, relativistic beaming in jets cannot be dominant. Instead, isotropic high-energy emission from the expanding lobes was predicted by~\cite{2008ApJ...680..911S} from studying the evolution of ultra-relativistic electrons injected from a terminal hotspot. Inverse-Compton scattering of seed photons by relativistic electrons in the jets was also proposed as a possible mechanism of non-thermal high-energy emission~\cite{2014ApJ...780..165M}. On~the other hand, a~thermal origin, free-free emission from the shocked plasma may also lead to $\gamma$-ray emission in CSOs~\cite{2009MNRAS.395L..43K}.
Among the few confirmed $\gamma$-ray CSOs, PKS\,1718$-$649 at low redshift ($z=0.014$) and with only $\sim 100$\,yr kinematic age may produce $\gamma$-rays via inverse-Compton scattering of the ambient photon fields by non-thermal electrons in the radio lobe~\cite{2016ApJ...821L..31M}. In~case of NGC\,3894 ($z=0.011$), another very young CSO, the~jet origin cannot be excluded~\cite{2020A&A...635A.185P}. Owing to the small number of such sources known to date, secure identification of any additional $\gamma$-ray-emitting CSO or larger medium-sized symmetric object (MSO) could provide a valuable test for models of (non-beamed) $\gamma$-ray emission in~AGN. 

The quasar CTD\,135 (2234 + 282, J2236 + 2828) emerged as a promising candidate, as~it is associated with a \textit{Fermi} source (4FGL J2236.3 + 2828,~\cite{2020ApJS..247...33A}), its radio morphology observed with cm-wavelength VLBI imaging at multiple frequencies is reminiscent of a CSO, and~its radio spectrum based on total flux density measurements appears to peak at around 1\,GHz~\cite{2016AN....337...65A}. However, determining the shape of the radio spectrum could be uncertain if the source is highly variable in time and the flux density measurements are not simultaneous. The~spectroscopic redshift of CTD\,135 is $z=0.790$. Its optical spectrum is characteristic of blazars in a bright continuum-dominated state, but~a broad Mg\,\textsc{II} emission line with observed-frame equivalent width of $4.9$\,\AA\, was detected in the spectrum in a low state~\cite{2012ApJ...748...49S}. Therefore the optical spectrum is transitional between flat-spectrum radio quasars (FSRQs) and BL Lac objects, affected by flux variability~\cite{2012ApJ...748...49S}.

The peaked radio spectrum, the~nearly symmetric, compact sub-kpc morphology, the~absence of Doppler-boosted radio emission, and~the slowly advancing component motion in CSOs~\cite{1982A&A...106...21P} are consistent with misaligned, young jetted radio sources. In~sharp contrast, blazars typically have flat or inverted radio spectra, a one-sided core--jet structure, a~compact core with high brightness temperature, indicating Doppler boosting in a jet closely aligned with the line of sight, and~often apparent superluminal expansion~speed.

However, relativistically beamed radio AGN could sometimes be confused with CSOs, if~structural information is considered only. A~typical core--jet morphology with a relatively bright jet component could mimic a compact double source~\cite{2021AN....342.1185R}. For~a more secure CSO classification, it is desirable to apply further selection criteria in addition to the classical ones, by~investigating radio flux density variability, and~the apparent separation speed between emission features seen in VLBI images with milliarcsec (mas) angular resolution~\cite{2021AN....342.1185R}. In~general, sources with beamed jets are expected to show rapid and high-amplitude flux density variability, as~opposed to misaligned radio AGN. The~apparent advance speed of CSO hotspots is generally lower than the core--jet separation speed derived from the jet component proper motions in blazars. Both  criteria above require a long series of radio observations, either flux density monitoring, or~multi-epoch VLBI imaging at a given~frequency. 

Another potential diagnostic tool is offered by the precise optical astrometry enabled by the  \textit{Gaia} space mission~\cite{2016A&A...595A...1G}. The~positional accuracy of \textit{Gaia} is comparable to that available with VLBI observations in radio. Most of the optical emission from an AGN is associated with the accretion disk surrounding the central black hole, perhaps with some contribution from the inner jet~\cite{2017A&A...598L...1K}, but~certainly not with the radio lobes. Therefore, a~compact double radio source without a clearly detected central VLBI core, but~with the \textit{Gaia} optical position available and falling in between the two opposite hotspots, is a strong indication for a CSO~\cite{2020MNRAS.496.1811K}.

A recent publication~\cite{2021RAA....21..201G} gave a comprehensive analysis of the observed properties, in~particular the $\gamma$-ray variability and broad-band spectral energy distribution (SED), of~CTD\,135. The~authors assumed that the source is a typical CSO, but~many findings of the study suggest that its properties are more consistent with a blazar.
This contradiction motivated us to revisit the case of CTD\,135 as a potential CSO candidate. In~Section~\ref{observations}, we collect multi-band observational data relevant for the source classification. In~the light of new observational evidence from multiple literature sources, we argue in Section~\ref{discussion} that the source is not a~CSO.
 

\section{Observational~Data}
\label{observations}
\unskip

\subsection{High-Frequency VLBI~Imaging}

The primary classification criteria for a CSO are the presence of compact radio emission within a projected linear size of $1$\,kpc, and~a symmetric compact double-lobed structure seen on both sides of a central source. Archival VLBI images at 2.3, 5, 8.4, and~15\,GHz analyzed by~\cite{2016AN....337...65A} indicated a barely resolved source in the northeast--southwest (NE--SW) direction. Circular Gaussian brightness distribution model components fitted to the highest-resolution 15-GHz data suggested three distinct components. However, the~central component, the~putative core, was blended with the NE feature in all other images made at frequencies lower than 15\,GHz. 

\textls[-25]{At that time, no VLBI imaging data at higher frequencies, and~thus with higher angular resolution, were available. The~15-GHz data used by~\cite{2016AN....337...65A} were obtained in the framework of the Monitoring of Jets in Active Galactic Nuclei with VLBA Experiments (MOJAVE) survey programme~\cite{2018ApJS..234...12L} performed with the U.S. Very Long Baseline Array (VLBA). CTD\,135 was monitored in 25 epochs from 1995 to 2015 and the data are available in the survey website (\url{http://www.physics.purdue.edu/astro/MOJAVE/sourcepages/2234+282.shtml}, accessed: 2 February 2022). In~a jet kinematic analysis of the first 19 years of 15-GHz MOJAVE observations,~\cite{2016AJ....152...12L} interpreted the mas-scale source structure differently from~\cite{2016AN....337...65A}, and~assumed the NE component as the core of a blazar with a one-sided jet pointing to SW. Based on measurements at 7 epochs,~\cite{2016AJ....152...12L} estimated the apparent jet component proper motion as $3.26\pm0.30$ times the speed of light ($c$), i.e.,~a superluminal value typical for blazars. In~turn, the~most recent jet kinematic analysis of the MOJAVE data~\cite{2021ApJ...923...30L} identified the core with the SW component, and~estimated almost negligible proper motion, $(0.58\pm0.53)\,c$, from~8 epochs of VLBA data. The~controversial proper motion estimates are the sign of the difficulties in reliably identifying jet features with less than $1$\,mas separation from the core in this source across the observing epochs at 15\,GHz. Multi-epoch MOJAVE 15-GHz polarization measurements~\cite{2018ApJ...862..151H} indicate that the fractional linear polarization is typically higher (up to $\sim 10\%$) in the NE component and lower or undetected in the SW component of CTA\,135, suggesting the latter is the weakly polarized core.}

\begin{figure}[H]

 \includegraphics[width=0.7\textwidth, bb=30 0 470 445, clip=, angle=-90]{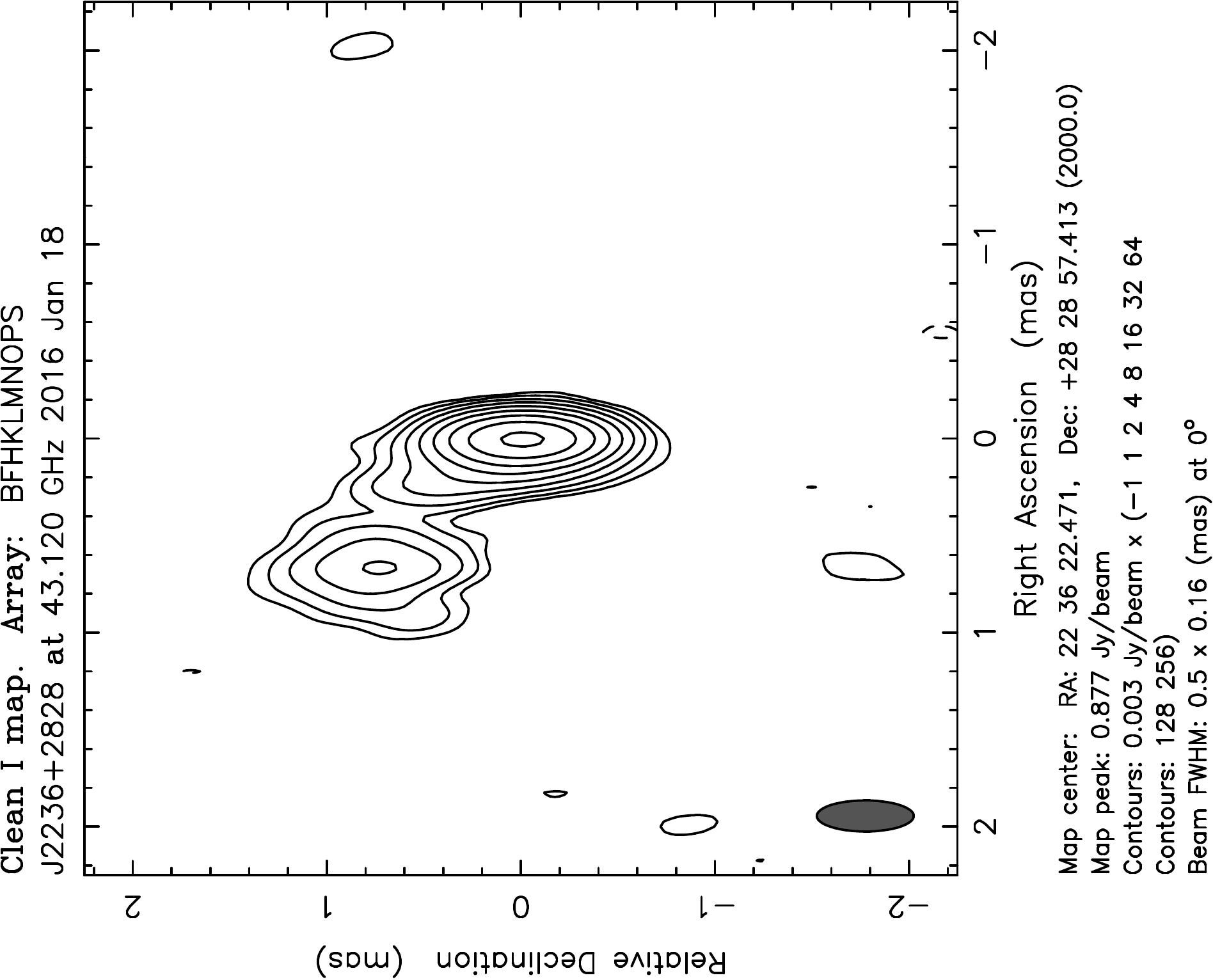}
\caption{The 43-GHz image of CTD\,135 based on VLBA data obtained on  18 January 2016~\cite{2020ApJS..247...57C}. The~peak brightness is 877~mJy\,beam$^{-1}$, the~lowest contours are drawn at $\pm 3$~mJy\,beam$^{-1}$ and the~positive contour levels increase by a factor of 2. The~half-power width of the elliptical Gaussian restoring beam is $0.5\,\mathrm{mas} \times 0.16\,\mathrm{mas}$ with the major axis oriented in the north--south direction, as~displayed in the bottom left~corner.
\label{q}}
\end{figure}

\begin{figure}[H]

\includegraphics[width=0.7\textwidth]{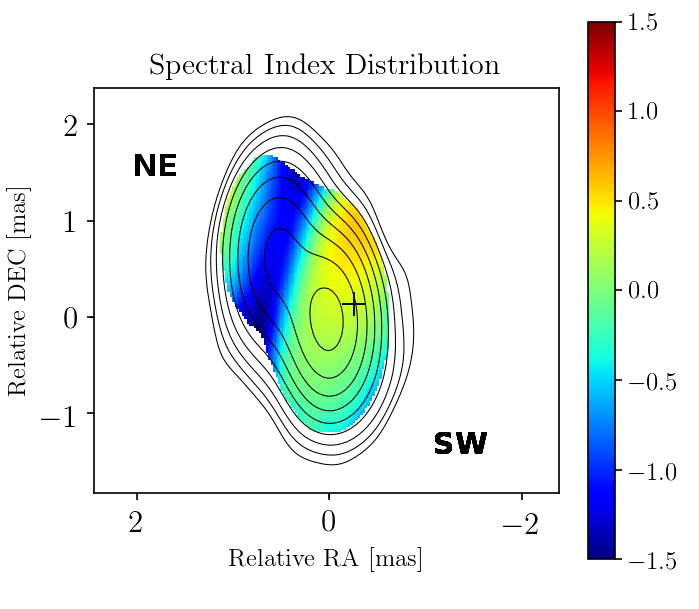}
\caption{Spectral index map of CTD\,135 between 15 and 43\,GHz frequencies, based on VLBA images obtained nearly simultaneously, within~1 month in 2015 December and 2016 January. The~half-power width of the common elliptical Gaussian restoring beam is $1.0\,\mathrm{mas} \times 0.5\,\mathrm{mas}$ with the major axis oriented in the north--south direction. The~contours, starting at 2~mJy\,beam$^{-1}$ and increasing by a factor of 2, show the 15-GHz brightness distribution. The~peak intensity is 726~mJy\,beam$^{-1}$. The~spectral index values are indicated by colours with the scale displayed in the right. The~cross marks the relative position of the \textit{Gaia} optical source, its size indicates the 1-$\sigma$ formal~uncertainties.\label{spx}}
\end{figure}

The unambiguous identification of the core with the SW component was made possible by the 43-GHz VLBA image published by~\cite{2020ApJS..247...57C}, as~CTD\,135 was included in the sample of 124 radio AGN observed in their survey. A naturally weighted image prepared from the same data in \textsc{difmap}~\cite{1994BAAS...26..987S} is displayed in Figure~\ref{q}. We constructed the spectral index image of the source and show it in Figure~\ref{spx}. It is based on the 43-GHz data obtained on  18 January 2016~\cite{2020ApJS..247...57C} and the 15-GHz MOJAVE data from  19 December 2015. The~epoch of the latter was chosen to be the closest in time to that of the 43-GHz image, to~minimize the effect of possible variability. The~colour scale indicates the spatial distribution of the two-point spectral index $\alpha$; it is defined following the convention $S \propto \nu^{\alpha}$, where $S$ is the brightness of the given image pixel and $\nu$ the observing frequency. The~spectral index image was constructed using the \textsc{vimap}~\cite{2014JKAS...47..195K} tool that allows us to properly align the images made at two different~frequencies.

\unskip   

Both the 15- and 43-GHz images were convolved  with a common restoring beam, using visibility data in the same ranges of VLBA baselines expressed in the units of wavelength. Both naturally weighted images were sampled with the same cell size ($0.025$\,mas). The~on-source integration times were $\sim54$\,min and $\sim20$\,min at 15 and 43\,GHz, respectively. The~typical error of the spectral index in the central regions of the map is estimated as $0.1$, considering also the flux density calibration uncertainties.
It is clear from Figure~\ref{spx} that the SW component has a flat spectrum ($\alpha \approx 0$), characteristic of a synchrotron self-absorbed core,  while the NE component has a steep spectrum (\mbox{$\alpha < -0.5$}) expected for an optically thin jet feature. This is evidence for a blazar with a core--jet~structure.

Another way to assess the spectral properties of the source is to characterize the brightness distribution of the core and the jet by fitting circular Gaussian components to the visibility data, and~calculate the two-point spectral index from the component flux densities at 15 and 43\,GHz. The~fitted core (SW) flux density is $S_\textrm{core}^{15} = 755 \pm 39$\,mJy and $S_\textrm{core}^{43} = 925 \pm 48$\,mJy at 15 and 43\,GHz, respectively. The~flux densities obtained for the jet component (NE) are $S_\textrm{jet}^{15} = 422 \pm 33$\,mJy and $S_\textrm{jet}^{43} = 178 \pm 15$\,mJy. Note that~\cite{2020ApJS..247...57C} fitted a core and two jet components to the same data but our choice of the simpler model is dictated by the 15-GHz data set where the additional inner jet component remains unresolved. The~two-point spectral indices are $\alpha_\textrm{core} = 0.20 \pm 0.10$ and \mbox{$\alpha_\textrm{jet} = -0.84 \pm 0.16$}.

The brightness temperature estimated for the core (i.e., the SW feature) at 43\,GHz is $T_B \gtrapprox 4 \times 10^{11}$\,K~\cite{2020ApJS..247...57C}. This exceeds the equipartition value~\cite{1994ApJ...426...51R} by almost an order of magnitude, indicating a relativistically beamed jet with a Doppler factor $\delta \gtrapprox 9-10$. Again, this is not expected for a CSO with a misaligned~jet.

\subsection{Spectral Energy~Distribution}

The broad-band SED constructed for CTD\,135 by~\cite{2021RAA....21..201G} using literature data spanning multiple electromagnetic wavebands from radio to $\gamma$-rays can be well-fitted with the two-zone leptonic radiation model usually applied for blazars, assuming a small jet inclination angle. Remarkably,~\cite{2021RAA....21..201G} estimate a Doppler factor $\delta = 10.8$ for the core region. This is consistent with the value deduced from the 43-GHz VLBI observations~\cite{2020ApJS..247...57C}. For~their SED fit,~\cite{2021RAA....21..201G} assumed a compact core component and a homogeneous extended (5\,mas) spherical emission region. Instead of the CSO, they suggest that the extended $\gamma$-ray structure is attributed to symmetric bubbles surrounding the~AGN.

\subsection{Radio~Variability}

As noted already by~\cite{2016AN....337...65A}, long-term non-simultaneous total flux density measurements of CTD\,135, as~well as archival VLBI flux density measurements, indicate significant variability at all frequencies above $\sim 1$\,GHz. They attributed this to the variability of the putative central core component, of~which the existence, however, is not confirmed by the new 43-GHz VLBI imaging~\cite{2020ApJS..247...57C}. The~core is highly self-absorbed and relatively weak at frequencies below 15\,GHz. It is therefore possible that the radio flux density variability is mostly associated with the NE (jet) component at lower frequencies and the SW (core) component at high~frequencies.

In their study on CSO identification problems,~\cite{2021AN....342.1185R} also mention CTD\,135 as an object exhibiting blazar-like flux density variability. By~accepting that this is a beamed core--jet source (Figure~\ref{spx}), the~variability properties can indeed be naturally~explained.

\subsection{$\gamma$-ray~Variability}

CTD\,135 is variable not only in the radio but in $\gamma$-rays as well, according to $\sim 11$\,yr of \textit{Fermi}/LAT measurements thoroughly analysed by~\cite{2021RAA....21..201G}. The~authors label the source as extremely variable in the $\gamma$-ray band, and~even claim the detection of a $\sim 460$-d quasi-periodic oscillation (QPO) in the light curve of CTD\,135. The~variability, together with the high $\gamma$-ray luminosity, indicate Doppler-boosted radiation~\cite{2021RAA....21..201G}, a~characteristic signature of blazars but not CSOs.
The $\gamma$-ray flux may have two origins. On~the one hand, a synchrotron self-Compton core component is responsible for the $\gamma$-ray QPO and the high-frequency radio variability. The~other contribution is a shocked component due to the interaction of the jet with the interstellar medium in the NE feature. The~latter would explain the ‘extended’ $\gamma$-ray emission component included in the SED fit~\cite{2021RAA....21..201G}.

\subsection{Infrared Colours and~Variability}

CTD\,135 was detected by the {\it Wide-field Infrared Survey Explorer (WISE)}~\cite{2010AJ....140.1868W} in all four of its wavebands, at~$3.4\,\mu$m (W1), $4.6\,\mu$m (W2), $12\,\mu$m, and~$22\,\mu$m (W4)~\cite{2014yCat.2328....0C}. Its infrared colours place it in the $\gamma$-ray blazar strip on the {\it WISE} colour--colour diagram~\cite{2011ApJ...740L..48M}. Additionally, it is listed as a likely blazar in the second {\it WISE} Blazar-like Radio-Loud Sources (WIBRaLS2) catalogue~\cite{2019ApJS..242....4D}.

The {\it WISE} satellite, after~the completion of its original mission, continued its all-sky survey in the framework of the NEOWISE (Near-Earth Object WISE) project~\cite{2014ApJ...792...30M} and afterwards as the NEOWISE Reactivation mission. Because~of the depleted cooling material, in~the prolonged projects the satellite remained operational only in the two shortest-wavelength (mid-infrared) bands, W1 and~W2.

\begin{figure}[H]
    
    \includegraphics[angle=-90, width=0.75\textwidth]{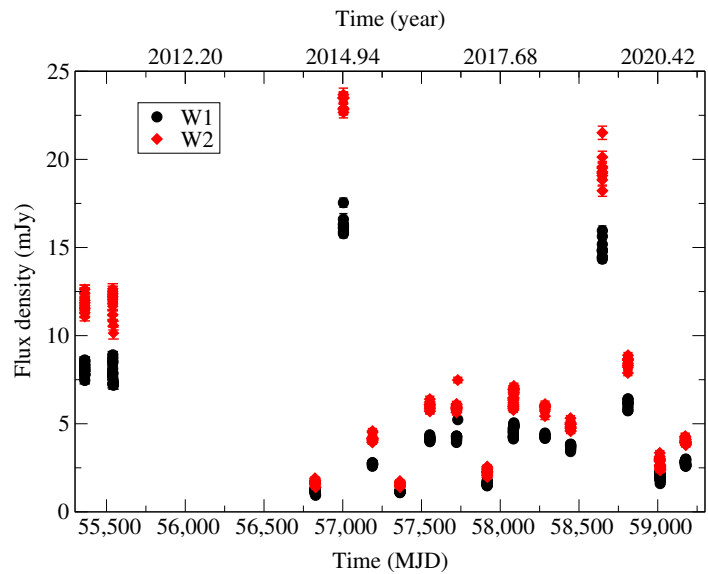}
    \caption{{\it WISE} light curve of CTD\,135 at $3.4\,\mu$m (W1) and $4.6\,\mu$m (W2), with~black and red symbols, respectively. }
    \label{fig:wise_lc}
\end{figure}

\begin{figure}[H]
    
    \includegraphics[angle=-90, width=0.75\textwidth]{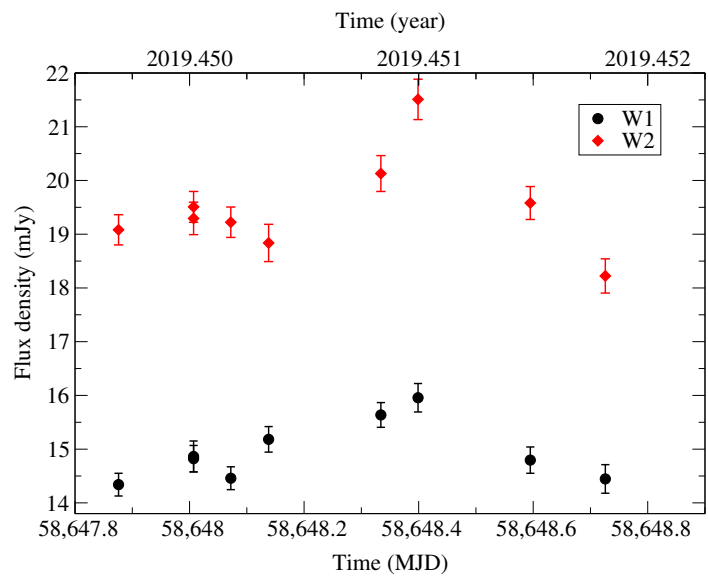}
    \caption{A zoom-in to the {\it WISE} light curve of CTD\,135 at the mission phase 13 in Figure~\ref{fig:wise_lc} at $3.4\,\mu$m (W1) and $4.6\,\mu$m (W2), with~black and red symbols, respectively.}
    \label{fig:wise13_lc}
\end{figure}

We downloaded the {\it WISE} single exposure data (\url{http://irsa.ipac.caltech.edu/Missions/wise.html}, accessed: 2 February 2022). Following the instructions of the Exploratory Supplement Series (\url{https://wise2.ipac.caltech.edu/docs/release/allwise/expsup/sec3_2.html}, accessed: 2 February 2022), we discarded measurements of bad quality. These involved data ($25$ measurement points) that have a profile-fit goodness value of $\geq3$ in either the W1 or W2 bands, indicating that a point-source model was not able to adequately describe the data (e.g., an image artifact was affecting the measurements). Then, we converted the magnitude values to flux densities and constructed the light curves shown in Figures~\ref{fig:wise_lc} and \ref{fig:wise13_lc}.

\unskip

CTD\,135 shows significant variation in its long-term infrared light curve in both bands (Figure~\ref{fig:wise_lc}). Notably, the~two epochs when its flux density reached the highest values, in~December 2014, near~Modified Julian Date (MJD) 57,000, and~in June 2019  ($\mathrm{MJD} \approx 58,650$), coincide with epochs of $\gamma$-ray brightening according to the light curve taken by the {\it Fermi} satellite~\cite{2021RAA....21..201G}.

In some mission phases, significant variability within a time span as short as a few days can also be seen in the infrared bands. An~example is shown in Figure~\ref{fig:wise13_lc}. Through causality argument, flux density change on a short time-scale implies small size for the emitting region responsible for the variability (e.g.,~\cite{2012ApJ...759L..31J}). Thus, the~variable infrared emission cannot originate from extended regions of the dusty torus or the host galaxy of CTD\,135. A coherent outburst lasting for $\sim 0.5$\,d in the observed frame (Figure~\ref{fig:wise13_lc}), if~corrected for the cosmological time dilation, implies an emission region within $0.5/(1+z)$ light days, or~$\sim0.0002$\,pc. A statistical study of the short-term infrared variability of {\it Fermi}-detected blazars by~\cite{2020MNRAS.494..764A} clearly showed the jet origin of their infrared emission observed in the {\it WISE}~bands.

In summary, the~infrared characteristics (colour and variability) of CTD\,135 make it similar to the beamed blazar~sources.

\subsection{VLBI and \textit{Gaia} Positions}

The accurate equatorial coordinates of CTD\,135 in the radio are right ascension $\mathrm{RA_{VLBI}} = 22^{\mathrm{h}} 36^{\mathrm{min}} 22.4708467^{\mathrm{s}}$ and declination $\mathrm{DEC_{VLBI}} = 28^{\circ} 28^{\prime} 57.412755 ^{\prime\prime}$ in the catalogue of the 3rd realization of the VLBI-derived International Celestial Reference Frame~\cite{2020A&A...644A.159C}. Here we consider the 32-GHz position (X/Ka-band solution (\url{https://hpiers.obspm.fr/icrs-pc/newwww/icrf/icrf3xka.txt}, accessed: 2 February 2022)) because we use high-frequency images for the spectral index map (Figure~\ref{spx}). The~formal errors are $0.07$\,mas and $0.08$\,mas for RA and DEC, respectively. In~the optical, the~most recent \textit{Gaia} Early Data Release 3 (EDR3)~\cite{2021A&A...649A...1G} lists (\url{https://gea.esac.esa.int/archive/}, accessed: 2 February 2022) the object with
right ascension $\mathrm{RA_{Gaia}} = 22^{\mathrm{h}} 36^{\mathrm{min}} 22.4708284^{\mathrm{s}}$ and declination $\mathrm{DEC_{Gaia}} = 28^{\circ} 28^{\prime} 57.412946^{\prime\prime}$ (J2000), with~$0.1$\,mas formal uncertainty in both coordinates and insignificant astrometric excess noise. The~radio--optical coordinate differences, $\Delta\mathrm{RA} = (0.31 \pm 0.12)$\,mas and $\Delta\mathrm{DEC} = (-0.19 \pm 0.13)$\,mas, are quite small, well within the values generally found between ICRF3 and \textit{Gaia} EDR3~\cite{2021A&A...652A..87L}. The optical position with the above error bars is marked with a cross in Figure~\ref{spx}. Although~it is very close to the brightness peak in the SW core, given the small overall angular extent of this radio source ($<$2\,mas), the~astrometric test is rather inconclusive regarding the CSO versus blazar classification. Especially because radio--optical positional offsets of up to $\sim 1$\,mas are ubiquitous in radio AGN, typically along the pc-scale jet direction~\cite{2017A&A...598L...1K}.      


\section{Summary---The Classification of CTD\,135 as a~Blazar}
\label{discussion}

In their recent publication,~\cite{2021AN....342.1185R} used CTD\,135 as one of the examples for potential 
CSO candidates, based on its radio spectrum composed of non-simultaneous flux density measurements, indicating strong variability. Indeed, as~it is suggested by independent observational evidence collected in Section~\ref{observations}, CTD\,135 shares its properties with blazars, AGN with relativistic jets pointed towards the observer. In~particular, the~existence of the compact flat/inverted-spectrum SW component (Figure~\ref{spx}), clearly resolved with VLBI at 43\,GHz and characterised by high brightness temperature~\cite{2020ApJS..247...57C}, the~highly variable $\gamma$-ray and mid-infrared emission, and~the broad-band SED indicating Doppler-boosted emission~\cite{2021RAA....21..201G} are all against the CSO classification tentatively proposed by~\cite{2016AN....337...65A}. 

We find it important to point out that CTD\,135 did not pass the scrutiny of a detailed multi-band investigation as a CSO candidate, and~it is in fact \textit{not} a CSO. While being labeled as a ``CSO candidate'', this AGN unfortunately already started its doubtful career in the literature as a (misclassified) rare $\gamma$-ray-emitting radio AGN with misaligned compact symmetric jets~\cite{2017ApJ...838..139S,2021A&ARv..29....3O,2021RAA....21..201G}. Following our study based on essential new observational data, we suggest removing CTD\,135 from the list of compact symmetric~objects.

\textls[-25]{Because of the difficulties in finding young $\gamma$-ray-emitting AGN~\cite{2016AN....337...59D}, it is of high importance to verify the CSO nature of each new candidate proposed. The~multi-waveband diagnostic methods applied here for the case of CTD\,135 could be useful for detailed studies of other potential $\gamma$-ray CSOs as well. The investigations of \textit{WISE} infrared variability and the radio--optical positional offsets based on VLBI and \textit{Gaia} measurements  in particular could offer a novel approach to this work.}



\vspace{6pt}
\authorcontributions{Conceptualization, S.F.; methodology, S.F., K.\'{E}.G., T.A.; writing---original draft preparation, S.F.; writing---review and editing, K.\'{E}.G., T.A.; visualization, K.\'{E}.G. All authors have read and agreed to the published version of the~manuscript.}

\funding{This research was funded by the Hungarian National Research, Development and Innovation Office (NKFIH), grant numbers OTKA K134213 and 2018-2.1.14-T\'ET-CN-2018-00001.}

\dataavailability{Calibrated VLBI data used for creating the images are available from the Astrogeo Center database (\url{http://www.astrogeo.org}, accessed: 2 February 2022). \textit{WISE} data used for creating the light curve are available from the NASA/IPAC Infrared Science Archive (\url{https://irsa.ipac.caltech.edu/Missions/wise.html}, accessed: 2 February 2022).} 

\acknowledgments{\textls[-25]{We thank the four anonymous referees for their comments and~suggestions. This research has made use of data from the MOJAVE database that is maintained by the MOJAVE team~\cite{2018ApJS..234...12L}. We acknowledge the use of archival calibrated VLBI data from the Astrogeo Center database maintained by Leonid Petrov. This publication makes use of data products from the \textit{WISE}, which is a joint project of the University of California, Los Angeles, and~the Jet Propulsion Laboratory/California Institute of Technology, funded by the National Aeronautics and Space Administration.}}

\conflictsofinterest{The authors declare no conflict of~interest.}

\newpage

\abbreviations{Abbreviations}{
The following abbreviations are used in this manuscript:\\

\noindent 
\begin{tabular}{@{}ll}
CSO & compact symmetric object\\
EDR3 & \textit{Gaia} Early Data Release 3\\
FSRQ & flat-spectrum radio quasar\\
LAT & Large Area Telescope (of the \textit{Fermi} satellite)\\
mas & milliarcsecond\\
MJD & Modified Julian Date\\
MSO & medium-sized symmetric object\\
MOJAVE & Monitoring of Jets in Active Galactic Nuclei with VLBA Experiments\\
QPO & quasi-periodic oscillation\\
SED & spectral energy distribution\\
VLBA & Very Long Baseline Array\\
VLBI & very long baseline interferometry\\
WISE & Wide-field Infrared Survey Explorer
\end{tabular}}


\begin{adjustwidth}{-\extralength}{0cm}

\reftitle{References}



\end{adjustwidth}
\end{document}